# Multi-wavelength Raman scattering of nanostructured Al-doped zinc oxide

V. Russo[1], M. Ghidelli[1], P. Gondoni[1], C. S. Casari[1,2], A. Li Bassi[1,2]

[1]*Dipartimento di Energia and NEMAS – Center for NanoEngineered Materials and Surfaces, Politecnico di Milano, via Ponzio 34/3, I-20133 Milano, Italy*
[2]*Center for Nano Science and Technology @PoliMI, Istituto Italiano di Tecnologia, Via Pascoli 70/3 I-20133 Milano, Italy*

In this work we present a detailed Raman scattering investigation of zinc oxide and aluminum-doped zinc oxide (AZO) films characterized by a variety of nanoscale structure and morphology and synthesized by pulsed laser deposition (PLD) under different oxygen pressure conditions. The comparison of Raman data for pure ZnO and AZO films with similar morphology at the nano/mesoscale allows to investigate the relation between Raman features (peak or band positions, width, relative intensity) and material properties such as local structural order, stoichiometry and doping. Moreover Raman measurements with three different excitation lines (532, 457 and 325 nm) point out a strong correlation between vibrational and electronic properties. This observation confirms the relevance of a multi-wavelength Raman investigation to obtain a complete structural characterization of advanced doped oxide materials.

## I. INTRODUCTION

Aluminum-doped zinc oxide (AZO), as an innovative indium-free transparent conductive oxide (TCO), shows appealing properties for several applications such as optoelectronic devices, energy efficient windows, low emissivity coatings and new generation photovoltaic devices.[1-5] Particularly in photovoltaic applications the interest lies both in films, for use as electrodes in 3rd generation solar cells, and in advanced nanoengineered structures in view of developing highly scattering layers or photoanodes for solar hybrid devices.[6,7] For all these applications a thorough comprehension of the material structure is required in order to understand and control the functional properties (e.g. transparency and electrical conduction), highlighting the role of dopant concentration, defects such as oxygen vacancies, and crystalline structure.

In this context Raman spectroscopy represents a powerful non-destructive characterization technique to study structural properties of complex oxides at a local level. It is based on the inelastic scattering of light, mostly in the visible range, by dynamic inhomogeneities of the system, e.g. phonons. Actually Raman scattering by phonons is largely determined by electrons that mediate the scattering event. Therefore Raman spectra contain information not only related to phonons, but also to electrons and electron-phonon interactions.[8,9] For this reason it has been largely used for semiconductor materials, such as zinc oxide, and for many other systems, for example carbon nanostructures and graphene,[10] where the investigation of electronic properties is fundamental. For the specific case of AZO we will show that a combination of Raman spectroscopy (RS) and Resonant Raman Spectroscopy (RRS) is able to provide a great amount of information both about structure, at the meso and nanoscale, and electronic properties, particularly related to carriers and doping.

In this paper we present a detailed characterization by multi-wavelength Raman spectroscopy of AZO films with a wide variety of nanoscale structure and morphologies at the sub-micrometer scale, from compact/columnar to porous hierarchical tree-like, deposited by Pulsed Laser Deposition and characterized electrically and optically in previous works.[7,11] In particular we focus on how the Raman spectrum of these systems is modified by doping, stoichiometry, carrier density and defects. To this aim a systematic comparison of Al-doped and undoped ZnO films with similar morphology is here presented and discussed, after a review of structural and vibrational properties of ZnO crystal and of its Raman spectrum in resonant and non-resonant condition (section III).

## II. EXPERIMENT

AZO films were grown on different substrates by ablating an $Al_2O_3$ (2 % wt.): ZnO sintered target with a ns-pulsed laser (Nd:YAG 4th harmonic, $\lambda$=266 nm, pulse repetition rate 10 Hz, pulse duration $\approx$ 6 ns). The laser fluence on the target was 1 $J/cm^2$, the target-to-substrate distance was fixed at 50 mm, while the background gas ($O_2$) pressure was varied from 0.01 Pa to 200 Pa. At low pressure (up to 2 Pa) a 500 nm thick film was obtained after 36 minutes (21600 pulses) of deposition, while for higher pressures the deposition time was adjusted in order to maintain a similar film thickness. Before deposition the substrates (soda-lime glass for XRD measurements, Ti for Raman spectroscopy) were cleaned in a RF plasma of Ar ions (accelerated by a 100 V potential) to improve film adhesion. Morphological characterization was performed by Scanning Electron Microscopy (Zeiss SUPRA 40 field-emission SEM). Structural characterization was conducted via X-Ray Diffraction (XRD) in theta-2theta configuration (Bruker D8 Advanced Diffractometer Cu $K_{\alpha 1}$, $\lambda$ = 0.15406 nm) and multi-wavelength Raman spectroscopy. Two different systems have been used for this characterization: the first was a Renishaw InVia micro-Raman spectrometer, equipped with an Argon laser emitting at 457 nm (in the following *blue* excitation) and an edge filter with cut-off at 150 $cm^{-1}$; the second was a Renishaw 2500 microscope system, including a frequency-doubled Nd:YAG laser emitting at 532 nm (*green* excitation) and a He-Cd laser, emitting at 325nm (*UV* excitation). The edge filter associated with the green excitation at 532 nm had cut-off at 50 $cm^{-1}$, allowing detection of the low-wavenumber ZnO mode (see below), while for UV excitation the cut-off was at 150 $cm^{-1}$. In all cases the power on the sample was kept below 1 mW to avoid laser-induced sample annealing and modifications.

## III. VIBRATIONAL STRUCTURE AND RAMAN SPECTRUM OF ZnO

Raman scattering is the inelastic scattering of photons by phonons. As direct photon-phonon coupling is weak for UV-visible photons, their interaction takes place mainly with the mediation of electrons. Indeed photons interact with electrons via an electron-radiation Hamiltonian, exciting the material into an intermediate (virtual) state, with the creation of an electron-hole pair. The e-h pair is then scattered into another state by emitting (or absorbing) a phonon, via the electron-phonon interaction Hamiltonian. Finally it recombines radiatively, emitting the scattered photon with a lower (higher) energy, while leaving the electronic state of the material unchanged.[8,9]

Therefore, as photons interact with the lattice via the electrons, Raman scattering measurements can provide information both on the vibrational structure and on the electronic properties of the material.[8]

First of all it is required to define the phonons that can be involved in the scattering, starting from the structure of the crystal and considering the fundamental Raman selection rule for which only phonons near $\Gamma$ (wavevector q ≈ 0) are measured.

At ambient pressure and temperature ZnO crystallizes in the wurtzite structure, a hexagonal lattice characterized by two interpenetrating sub-lattices of $Zn^{2+}$ and $O^{2-}$ ions, such that each Zn ion is surrounded by a tetrahedron of oxygen ions and vice versa. This arrangement gives rise to polar symmetry along the hexagonal vertical axis (c axis).

The wurtzite structure of ZnO crystal has $C_{6v}$ symmetry. There are 4 atoms in the hexagonal unit cell leading to 12 phonon branches, 9 optical and 3 acoustic, classified according to the following irreducible representations: $\Gamma = 2A_1 + 2B_1 + 2E_1 + 2E_2$, where A and B modes present one-fold degeneracy and E modes are twofold. One $A_1$ mode and one $E_1$ pair are the acoustic phonons. The 9 optical phonons are divided into one $A_1$ branch (both Raman and IR active), one doubly degenerate $E_1$ branch (both Raman and IR active), two doubly degenerate $E_2$ branches (Raman active only) and two inactive $B_1$ branches. Therefore there are 4 Raman active phonons at the centre of the Brillouin zone.[12] Complete phonon dispersion relations, calculated at zero pressure and temperature, are reported in ref.13 together with Inelastic Neutron Scattering experimental data.

Actually, the Raman spectrum of bulk ZnO presents 6 first-order peaks,[12,14-16] not 4 as the number of active modes, as visible in fig. 1 where the unpolarized spectrum of our bulk ZnO target is reported for reference, together with a sketch of the atomic displacements of the modes. The two most intense peaks are associated to the $E_2$ modes, the first at 99 cm$^{-1}$ (named $E_2^{low}$), dominated by the vibrations of the heavy Zn sub-lattice, the second at 437 cm$^{-1}$ (named $E_2^{high}$) which involves mostly the oxygen atoms. The atoms move perpendicular to the c-axis and, in each sublattice, the neighbouring ions move opposite to each other: hence the displacements of the ions sum up to zero, resulting in no net induced polarization. For this reason these two modes are defined non-polar modes. On the contrary the $A_1$ and $E_1$ phonons, that are oxygen-dominated, are polar modes. The atoms move parallel or perpendicular to the c axis for $A_1$ or $E_1$ symmetry, respectively, in such a way that the displacement of the Zn sublattice with respect to the oxygen sublattice induces a net oscillating polarization. As a consequence the two modes $A_1$ and $E_1$ each split into LO and TO components because of the polarity induced macroscopic electric field associated to the LO modes, which acts as an additional restoring force for the ion oscillation. For this reason the associated Raman peaks become 4, i.e. $A_1$(TO) at 380 cm$^{-1}$ and $A_1$(LO) at 574 cm$^{-1}$, $E_1$(TO) at 407 cm$^{-1}$ and $E_1$(LO) at 583 cm$^{-1}$, which together with the two $E_2$ account for the 6 detectable peaks (see fig.1 and table I for reference values and their variability range based on the literature).

Each of these Raman features can be enhanced or suppressed in polarized Raman scattering measurements with a suitable combination of crystalline orientations and polarization of the incident/scattered light, as reported in ref. 17. This property is significant for single crystal ZnO or in case of crystalline films with preferential growth orientation.

The spectrum is somehow complicated by the presence of some well resolved multi-phonon features, in particular one at about 200 cm$^{-1}$ (overtone of the $E_2^{low}$ mode) and one at about 330 cm$^{-1}$ (due to the combination $E_2^{high}$ - $E_2^{low}$), with comparable intensity and similar shape as first order peaks. Finally two broad second-order bands appear, one in the range 540-700 cm$^{-1}$, due to optical-acoustic combinations, the other in the range 980-1200 cm$^{-1}$ for optical combinations.[15]

As stated before the electron-phonon interaction plays a fundamental role in the Raman scattering. Two types of interaction must be considered, the *deformation potential interaction*, due to the modulation of the crystal periodic potential by the relative atomic displacements of the phonons, and the *Fröhlich interaction*. The latter is related to the macroscopic electric field generated by the relative displacement of oppositely charged atoms within the primitive cell and therefore concerns only LO phonons. For ZnO there are two LO modes, the $A_1$(LO) and the $E_1$(LO), with close wavenumbers (574 and 583 cm$^{-1}$). However visible Raman cross section of the $A_1$(LO) mode is nearly vanishing in ZnO, due to the virtual compensation of the two scattering terms cited above.[18] Moreover the two modes have different polarization selection rules in backscattering geometry, being the $A_1$(LO) allowed from the c face, while the $E_1$(LO) is a forbidden mode from both the a and the c faces. For all these reasons the two modes are often merged in a single band, called the LO band, especially in disordered materials where the polarization selection rules are relaxed. Some works interpret this band as a LO-quasimode with mixed $A_1$ and $E_1$ symmetry induced by the presence of defects, an effect that is enhanced using resonant conditions,[19] as explained in the following

Resonance Raman scattering (RRS) is a well-known phenomenon that occurs when the excitation energy is close to or larger than the optical gap. ZnO is a semiconductor with wide direct bandgap (3.3 eV) and large exciton energy (about 60 meV) that gives rise to a strong photoluminescence (PL) peak in the UV corresponding to band-to-band transitions.[16,20] Also broad PL bands in the visible range can occur for transitions involving deep levels inside the gap, which are related to defects or vacancies in the crystal structure.[21,22] If UV excitation energy (above 3 eV) is used for Resonant Raman measurements the spectrum appears to be dominated by the LO mode and its multiples eventually superimposed upon PL bands.[23] The strong enhancement of the solely LO modes by RRS can be understood in terms of the Fröhlich electron-phonon interaction, which is dominant near the band edge in polar semiconductors.[18] Interestingly also with lower excitation energy, even in the visible range, some *pre-resonance* effects appear. A pioneering study by Calleja et al.[24] investigated the variation of the intensity of the Raman peaks for different visible excitation energies, showing an increasing trend with increasing energy, mostly for the LO modes and for the second-order modes. More recent works studied RRS in doped thin films[25] and nanoparticles,[26] showing additional enhancement of the LO modes due to defect-induced RRS. Generally the most important aspect to underline is a variation with excitation energy of the relative intensity between $E_2$ peaks and the LO band, from a prevalence of the first using visible excitation to the opposite situation with UV excitation.

In the case of ZnO thin films the Raman spectrum appears similar to the case of bulk ZnO, as shown by several authors,[16,27,28] particularly if the film has been grown in presence of oxygen and/or keeping the substrate at high temperature (above 400°C). Anyway some differences must be highlighted, mostly relative to the LO band. For example Tsolov et al.,[29] reporting about ZnO films grown by magnetron sputtering, showed evident broadening and red-shift of the LO band with reduced oxygen partial pressure during deposition, combined with an increment of its relative intensity with respect to $E_2^{high}$ peak. They also showed an opposite trend upon annealing in air. Similar results are reported by many other works,[16,30] suggesting a non-trivial correlation between shape and intensity of the LO band and oxygen stoichiometry.

Further aspects must be accounted for when considering the case of aluminum-doped ZnO. Many works[28,29,31,32] investigated the role of extrinsic dopants such as Al on the Raman spectrum of doped ZnO. Generally they showed some modifications induced in the spectrum by dopants, resulting again in broadening and intensity increasing of the LO band. This aspect has been used to justify an analogy between extrinsic and intrinsic doping attributed to oxygen vacancies. Moreover, additional anomalous modes (AM) have been shown to appear in the Raman spectrum of doped-ZnO (for instance in the case of AZO[31] at 275 and 510 cm$^{-1}$), for which different explanations have been proposed and largely debated in the literature. A possible origin has been attributed to breakdown of the translational crystal symmetry induced by defects and impurities or local electric field at grain boundaries, with consequent activation of otherwise silent modes ($B_1$ modes[29,33,34]). On the other hand, completely new doping-related modes have been suggested, e.g. Local Vibrational Modes (LVMs), if related to molecular-like vibrations of impurities, doping and defect complexes,[35] or Surface Phonon Modes (SPMs), highly localized near grain boundaries.[31] Also the presence of hydrogen or nitrogen as unavoidable source of doping has been used to explain the presence of these anomalous modes.[34,36] All these explanations presents some level of controversy and appear to be mutually exclusive, leaving the subject as an open question

## IV. RESULTS

### A. Structure and morphology

The deposited AZO films show a variation in morphology with deposition pressure, moving from dense, compact films, characterized by oriented columnar structures with variable size at low pressure (up to 2Pa), to porous films constituted by nanoparticles assembled in hierarchical, tree-like structures at high pressure (above 10 Pa), as visible in SEM images reported in fig. 2. This behaviour is due to the interaction of the ablated species with the background gas during deposition, as discussed in detail in our previous works.[7,11] A similar variation in morphology with deposition pressure was observed for ZnO films (not shown), although in general they result slightly more compact than corresponding AZO films deposited at the same pressure.

From now on, for simplicity, we will refer to our films as *compact* films (deposited at or below 2 Pa) and *porous* films (deposited above 10 Pa), both for AZO and ZnO. Structural characterization performed by X-ray diffraction (not shown here, see ref. 11 for data relative to AZO films) highlighted a preferential growth direction along the c-axis of the wurtzitic hexagonal lattice, for both ZnO and AZO compact films. The position of the corresponding (002) peak was always lower than the ZnO equilibrium value ($2\theta_{(002)} = 34.43°$), moving towards such a value with increasing oxygen pressure, while the width of the peak decreased with increasing oxygen pressure.

In fig. 3 we report the variation of the lattice parameter c and of the vertical domain size (calculated from XRD spectra) with oxygen pressure, for both kinds of films. For AZO films a high value of the lattice parameter c is found for compact films, up to 5.34 Å at 0.01 Pa, with a decrease towards bulk value ($c_{bulk} = 5.20$ Å) with increasing pressure. The trend is similar for ZnO films, but with lower deviation from the bulk value. The explanation, according to ref. 37, lies in the non-stoichiometry of the compact films deposited at low oxygen pressure. The high oxygen vacancies concentration forces the $Zn^{2+}$ and $Al^{3+}$ ions to occupy lattice interstitial positions (rather than the equilibrium ones), thus leading to an increment of the vertical dimension of the wurtzitic cell, much more pronounced in the case of AZO for the presence of $Al^{3+}$ ions. Another contribution altering the equilibrium geometry of the wurtzitic cell can be represented by residual compressive stress at the film-substrate interface arising during the film growth.[11] Both effects contribute also to increase structural disorder. This prevents the formation of crystalline domains of relevant dimensions at very low pressure, and domain size (calculated with Scherrer's formula applied to the (002) peak) remains lower than 10 nm up to 1 Pa in the case of AZO films. Above this pressure a rapid increase of the domain size is evident with a maximum value at 2 Pa. Again ZnO films confirm to be less disordered than AZO showing domain sizes greater than 10 nm already at 0.01 Pa.

Moving to porous films deposited at higher pressures (above 2 Pa) the preferential growth direction along the c-axis of the ZnO wurtzitic structure is progressively lost and other reflexes at angles typical of polycrystalline ZnO appear in the XRD diffractograms, indicating a transition towards randomly oriented but stoichiometric crystals with reduced dimensions. Accordingly the domain size starts to decrease again (after 2 Pa) and porous AZO films can be depicted as a collection of nanoparticles with progressive saturation of oxygen vacancies. For ZnO films the reduction of the domain size appears at pressure higher than 100 Pa, confirming the additional contribution of Al to structural disorder of AZO films.

**B. Raman spectroscopy**

The comparison between Raman spectra of a selection of ZnO and AZO films deposited at the same oxygen pressure (namely at 0.1, 2, 10 and 100 Pa) is reported in figs. 4(a), 4(b) and 5 for the three used excitation wavelengths (532, 457 and 325 nm), respectively.

*1. Green excitation (532nm)*

Fig. 4(a) shows Raman spectra from ZnO and AZO films collected with the *green* excitation wavelength (532 nm, 2.3 eV), far from the resonance condition. As explained in the experimental section only with this configuration (edge filter cut-off at 50 cm$^{-1}$) it is possible for us to detect the $E_2^{low}$ mode expected at 99 cm$^{-1}$.

A common aspect relative to both ZnO and AZO is that at low oxygen deposition pressure (below 2 Pa) the spectra show two large bands, one at lower wavenumbers (below 300 cm$^{-1}$), mostly related to vibrations of the Zn sublattice, and the other at higher wavenumber (300-700 cm$^{-1}$), resulting from oxygen-related modes. The overall shape resembles the zinc oxide vibrational density of states (see fig. 1 in ref. 13), as expected for disordered materials.[38] Conversely, with increasing oxygen pressure (above 2 Pa) an evolution towards more ordered structures, at least locally, is attested from the appearance of peaks in positions very close to the crystalline peaks. This is also an indication of oxygen stoichiometry completion.

These general observations relative to oxygen vacancies and stoichiometry are in accordance with XRD analysis.[11] As explained above, in the case of compact films, the shift with decreasing pressure of the position of the dominant (002) peak with respect to bulk ZnO suggested deformations in the lattice due to the presence of oxygen vacancies, while the appearance of typical ZnO powder reflexes for increasing pressures indicates that porous films are constituted by stoichiometric nanoparticles with no preferential orientation.

Anyway some distinctions between Raman spectra of ZnO and AZO must be noted. First in the case of very low pressure (see the two lower spectra relative to deposition at 0.1 Pa in fig. 4(a)) the two bands (especially the high wavenumber one, related to oxygen) are larger and less structured for AZO than for ZnO, proving the additional disorder introduced by aluminum doping, as confirmed also by XRD results (see fig. 3(a)).

Moving to deposition at 2 Pa relevant differences between the two systems can be observed. As regards pure ZnO the spectrum starts to present some bulk features, namely the two $E_2$ peaks at about 99 and 437 cm$^{-1}$, emerging from an overall band-like spectrum. Nevertheless, the intensity ratio among Raman features is very different from the case of bulk, where $E_2$ peaks dominate (see fig. 1), being the LO bands much more intense than all other features. This aspect is probably related to an enhancement of the LO band via *pre-resonant* effects, due to the presence of defects that induce a red-shift of the resonant conditions (see discussion section). Even more intriguing is the case of AZO film deposited at 2 Pa. This film showed very interesting optical and conductivity properties (resistivity of about 4x10$^{-4}$ Ω cm, transparency greater than 80% in the visible range), as reported in our previous work,[7] with an optimal combination of carrier concentration, provided by shallow levels due to defects/doping, and good mobility ensured by the highest value of crystalline domain size (see fig. 3(b)). Its spectrum still presents an overall aspect typical of disordered materials, characterized by two large bands and no clear evidence of bulk-like features, but it also exhibits some peculiarities. First of all the LO band clearly dominates and appears significantly broad. Moreover two distinct and sharp peaks arise, one at 275 cm$^{-1}$ and the other at 510 cm$^{-1}$, not related to active modes of ZnO, but analogous to peaks cited in many works about Raman of AZO and other doped-ZnO films. Both can be defined as Anomalous Modes (AM), whose debated origin has been already discussed in section III. The probable cause of the appearance of these peaks only in the spectrum of the 2 Pa AZO film is still under investigation. Here, we note the sharpness of these peaks with respect to the rest of the spectrum, which suggests a molecular origin and in our opinion should rule out the hypothesis of activated silent modes.

In fig. 4(a) it is possible to note that for films deposited at higher oxygen pressure the differences between the two systems (AZO and ZnO) tend to become less marked with increasing pressure. For pure ZnO films spectra start to show all the six active modes, with the expected intensity ratios, at 10 Pa. At the same pressure also AZO spectra present all the 6 peaks, but with some residual disorder effects, such as broadening and anomalous intensity ratios between $E_2$ modes and LO band. Finally at 100 Pa spectra for both ZnO and AZO films are very similar to the spectrum of bulk ZnO, attesting local order and stoichiometry of the two systems. In particular, the evolution of the two $E_2$ modes with deposition pressure is similar for both ZnO and AZO films. The $E_2^{high}$ peak, mostly related to vibration of the oxygen sub-lattice, becomes more intense and sharper with increasing oxygen content starting from 10 Pa, while the peak $E_2^{low}$, which is dominated by vibrations of the zinc atoms, appears at 2 Pa and becomes evident from 10 Pa on, proving the formation of the zinc sub-lattice (see section V for a quantitative analysis relative to position, width and intensity). Only in the case of AZO films the $E_2^{low}$ peak appears to be superimposed to a large band extending towards very low wavenumbers (below the filter cut-off). Different explanations for the origin of this band may be suggested, possibly related to Al doping. Among these we cite for example distortions of the Zn sub lattice if aluminum atoms randomly occupy interstitial positions with consequent recurrence of the low wavenumber band; light scattering by coupled longitudinal plasmon-phonon modes, related to carriers provided by doping;[7,39] tail of visible photoluminescence due to defect/doping levels inside the gap.[21,22]

### *2. Blue excitation (457 nm)*

Moving to *blue* excitation wavelength (457 nm, 2.71 eV), still an off-resonant condition for bulk ZnO, the general evolution of the spectra with deposition pressure is similar to the case of *green* excitation described above, with the two large bands at low oxygen pressures and bulk-like peaks at higher pressures, see fig. 4(b) (note the filter cut-off at 150 cm$^{-1}$ hiding the $E_2^{low}$ mode). The most important difference with respect to *green* excitation is a strong enhancement of the LO band for all deposition pressures. In the case of compact films the LO band completely dominates the spectrum, with a smoothing of the other features, while in the spectra of porous films it increases up to show comparable intensity with respect to the $E_2^{high}$ mode. This effect agrees with ref. 24 predicting increase of LO mode intensity with increasing excitation energy in bulk ZnO. Here, a further enhancement can be inferred, particularly in the case of more disordered systems, suggesting an additional defect-induced effect, via the variation of the electronic properties of the material (see discussion).

Interestingly also the AM at 275 cm$^{-1}$ in the spectrum of the 2Pa AZO film seems to be enhanced with *blue* excitation, emerging even in the case of AZO films deposited at 0.1 Pa (see the lowest red spectrum in fig. 4(b)). The dependence of this peak from excitation wavelength indicates that its unclear origin may be related to electronic properties of the material, specifically to the doping-related ones.

### *3. UV excitation (325 nm)*

Resonant spectra collected with *UV* excitation (325 nm, 3.81 eV) are expected to be completely different compared to the case of visible excitation, with the presence of the LO mode only and its overtones at multiple wavenumbers.[23] In our case (see fig. 5) the spectra of pure ZnO films reproduce this expectation for all pressures, showing also an increasing background related to a PL band in the visible range, which suggests the presence of defect levels in the gap. The PL contribution is more pronounced (relative to LO peaks) for low pressure films, where the presence of defects such as oxygen vacancies is more likely. On the contrary for AZO films the typical resonant spectrum appears only at high pressures (above 10 Pa) while spectra for films deposited at low pressure do not show resonance effect, but rather a behaviour similar to the case of *blue* excitation (*pre-resonance* condition). In our previous work[7] we reported the estimation of the optical gap of these AZO films, finding very high values for compact films (above 3.8 eV), evidence of the Moss–Burstein effect,[40] and bulk values for porous films (3.3 eV), hence below the used UV excitation energy. This can justify the different resonant behaviour in the UV range for AZO films deposited at low and high oxygen pressure.

## V. DISCUSSION

A quantitative analysis of visible spectra collected with *green* excitation has been performed by fitting the Raman data in the range 50-700 cm$^{-1}$. The initial guess consisted of Lorentzian peaks in the number of the expected Raman features in this spectral range, six first order modes and two well-defined second order modes, leaving free all the parameters (position, width and area). Some care must be taken in the case of spectra dominated by bands, as for the low pressure films, and for an amorphous-like background due to residual disorder for the higher pressure films. Two free Gaussian functions have been added, one for each band of the disordered system.

We focus on the properties of the two $E_2$ modes, the LO band and the relation among them. We report as a function of pressure the variation of peak position and width for the two $E_2$ modes in fig. 6 and of the intensity ratio between the LO band and the $E_2^{high}$ mode (i.e. ratio of the area of the fitting curves) in fig. 7.

## A. $E_2^{low}$ mode

This is a non-polar mode dominated by the vibrations of the heavy Zn sub-lattice. A small variability of the peak position values with oxygen pressure has been found both for ZnO and AZO films (fig.6(a)). For AZO films the peak position is always very close to the bulk ZnO value, especially for pressure higher than 100Pa, where also FWHM is very small (about 6 cm$^{-1}$), attesting a high degree of local order in the Zn sublattice at any pressure. This suggests that the presence of aluminum atoms does not prevent the formation of the Zn sublattice, rather influencing the oxygen sublattice (see below). In the case of ZnO, excluding the 2 Pa film which shows the highest value (100 cm$^{-1}$), the peak position for ZnO films is around 94 cm$^{-1}$, lower than the bulk value, while its FWHM decreases almost linearly with increasing pressure (fig. 6(b)), though remaining larger than 10 cm$^{-1}$. Both these aspects can account for a residual stress in pure ZnO films at all pressures.

## B. $E_2^{high}$ mode

This non-polar mode is associated to vibrations of the oxygen sub-lattice. In ZnO films the peak position is nearly constant for all deposition pressures, but always lower than the bulk value (fig. 6(a)), in agreement with the hypothesis of residual stress, further confirmed by FWHM values larger than 10 cm$^{-1}$ (fig. 6(c)). Anyway, the small variation of the values of both position and width suggests that the oxygen sub-lattice is formed at relatively low pressure for pure ZnO films. On the contrary, a decreasing trend from quite high values (about 446 cm$^{-1}$ in compact films) towards the bulk value for pressure higher than 100 Pa is shown for AZO films, combined with a strong decreasing trend for FWHM from more than 40 cm$^{-1}$ at 2 Pa to about 10 cm$^{-1}$ at 100 Pa. These data indicate the presence of a significant local disorder in the oxygen sublattice in the case of AZO at low and intermediate pressure (up to about 100 Pa) and support the hypothesis of a correlation between stoichiometry defects and Al doping.[11,37] Finally, for pressure higher than 100 Pa the width of the peak starts to increase again, though not changing its position significantly. This can be due to effects related to confinement or surface/grain boundaries (see also discussion about LO band), more relevant in porous films which are constituted by an assembly of nanoparticles with decreasing size at increasing pressure.[7,11]

**C. LO band**

The two LO modes of ZnO are combined in a single LO band for disordered or polycrystalline material. Interpretation of this band is very complex due to many reasons: first, the mixed nature of the band derived from a combination of the $A_1(LO)$ and $E_1(LO)$ modes with different symmetries and different selection rules in crystalline ZnO; second, the polar character of the involved modes with consequent activation of the two scattering terms described in section III; third, the enhancement of the Frohlich electron-phonon interaction near resonance that enormously increases LO intensity; finally the prevalence of oxygen vibrations in the dynamics of the LO modes (see fig. 1), suggesting a correlation with the stoichiometry of the material.

In order to perform a comparison among the different ZnO and AZO films, we here focus on the ratio between the LO band and the $E_2^{high}$ peak intensity, indicated by $I(LO)/I(E_2^{high})$. In fig. 7(a) the variation of this parameter with deposition pressure is reported both for ZnO and AZO films in the case of *green* excitation, showing a general decreasing trend with increasing oxygen pressure. The reference bulk ZnO value (about 0.75), attesting the expected prevalence of the $E_2^{high}$ mode, is calculated from data in fig. 1 and is indicated in the graph with a horizontal dashed line. For ZnO films a strong prevalence of the LO band with respect to the $E_2^{high}$ mode is evident at very low pressure (intensity ratio much greater than 1), but the reference bulk value is rapidly reached (already at 2 Pa). Conversely for AZO films the $I(LO)/I(E_2^{high})$ ratio remains greater than the bulk value up to 50 Pa, reaching it at about 80 Pa. Finally, the ratio slightly increases again at the highest pressure (200 Pa) for both systems.

The interpretation of this behaviour is not straightforward, but some interesting remarks can be highlighted. First of all both $E_2^{high}$ mode and LO modes are oxygen-dominated vibrations, as explained in section III, but they present an opposite trend with the absolute content of oxygen in the material. Particularly in our films we observe an increase of the $E_2^{high}$ intensity and a decrease of the LO band with increasing oxygen deposition pressure, as evident from Raman spectra reported in fig. 4. The general tendency of their ratio with increasing oxygen is towards the bulk situation, where the non-polar $E_2^{high}$ mode largely dominates. Therefore very interestingly a value lower than 1 for the intensity ratio, combined with the position and width of the $E_2^{high}$ peak, could be used as a hint for oxygen stoichiometry. In our case the stoichiometry can be considered complete using oxygen pressure higher than 2 Pa for ZnO films and higher than 80 Pa for AZO films. This difference highlights an influence of doping on stoichiometry.

At very high pressure, where films are extremely porous, a further variation of the intensity ratio is noticed, and a value higher than 1 is obtained both for ZnO and for AZO. For these films a broadening of the peaks is observed (not shown here), but no significant variations in their positions (see fig. 6). In addition the XRD results discussed above (see fig. 3) have shown a decrease of the domain size at high pressure, suggesting a possible activation of effects related to confinement or surface/grain boundaries on Raman mechanisms. The net effect is the prevalence of the LO band again, although at a less extent with respect to the case of strongly sub-stoichiometric films. Excluding a significant reduction of the $E_2^{high}$ intensity, the increased intensity ratio can be attributed to LO enhancement via a *pre-resonance* effect.

This parameter varies significantly also with excitation wavelength, as visible in fig. 7(b), where the comparison between *green* and *blue* excitation in the case of AZO films is reported. This aspect is not surprising because it is known that LO mode intensity increases with increasing excitation energy even in bulk ZnO (see section III), while the $E_2$ mode does not. Consequently the examined ratio also varies. The general trend with deposition pressure is similar, but with higher values using the *blue* excitation, as expected.

Finally we note that the compact AZO films, showing very good conductive properties (in particular the one deposited at 2 Pa[7,11]), present very high values of the $I(LO)/I(E_2^{high})$ ratio both with *green* and *blue* excitation, due to both sub-stoichiometry and doping, suggesting that a high value of this parameter can be used as a hint of low resistivity.

From all these observations a general correlation is evident between the relative intensity of the LO mode and any kind of disorder, in particular oxygen vacancies, doping and the combined action of the two. The question is very complex, but we suggest an indirect influence of disorder on *pre-resonance* conditions via modification of electronic properties of the material (e.g. creation of defect levels inside the gap) with a consequent enhancement of the LO band. Consequently the $I(LO)/I(E_2^{high})$ ratio appears to be an extremely interesting parameter.

## IV. SUMMARY

In summary we presented a characterization of compact and nanostructured Al-doped ZnO films by multi-wavelength Raman Spectroscopy, obtained performing a systematic comparison of Raman data for nanostructured ZnO and AZO films with similar morphology. We found that any variation of both structural and electronic properties due to defects, doping, stoichiometry, confinement strongly influences widths, positions and intensities of Raman peaks and their variation with excitation wavelength. Therefore we summarize our results on the base of these effects.

*(a) Local structural order*. From the structural point of view the Zn sub-lattice appears to reach an ordered configuration at relatively low pressure both for ZnO and for AZO films, as attested by position and width of the $E_2^{low}$ peak. On the contrary, significant differences have been found related to the oxygen sub-lattice for the two systems. In fact, for ZnO the position and width of the $E_2^{high}$ mode and the intensity ratio $I(LO)/I(E_2^{high})$ tend towards bulk-like values at about 2 Pa with no further variation. Differently AZO films have shown a more disordered oxygen configuration, as revealed both by a wide variability of position and width of the $E_2^{high}$ mode with pressure and by an intensity ratio $I(LO)/I(E_2^{high})$ greater than one up to about 80 Pa. This suggests a strong influence of Al-doping, which appears to favour the presence of oxygen vacancies, giving rise to a disordered oxygen sub-lattice.

*(b) Oxygen stoichiometry.* Our Raman results shed light on the issue of oxygen stoichiometry, by using information extracted by two oxygen-related Raman features, the $E_2^{high}$ peak and the LO band. In particular, we have analyzed the position and width of the $E_2^{high}$ mode to study the local order of the oxygen sub-lattice, as just explained above, and the intensity ratio $I(LO)/I(E_2^{high})$, which remains well above bulk value of 0.75 if the film is sub-stoichiometric. By a suitable combination of these data it is possible to infer if the film has reached stoichiometry or not.

*(c) Doping.* Many variations in the Raman spectra of AZO films with respect to ZnO films, strictly depending on the presence of Al atoms, have been discussed. In particular we want to underline the suppression of the UV resonance conditions in the case of the conductive compact AZO films deposited at pressure up to 2 Pa. This aspect is an indirect confirmation of energy band-gap enlargement determined by doping via Moss–Burstein effect. Moreover, the appearance of new so-called Anomalous Modes in the visible Raman spectra is related to the presence of doping in sub-stoichiometric films, particularly the ones that have shown very good conductivity properties.

In conclusion, all the above observations suggest a strong correlation between Raman spectra and electronic properties of AZO films, indicating the relevance of a multi-wavelength investigation to obtain a complete structural and electronic characterization of this materials.

**TABLE CAPTIONS**

TABLE I. Raman modes of wurtzitic ZnO crystal. The range of frequency variability is based on literature data[12-16]

| Symmetry | Frequency (cm$^{-1}$) |
|---|---|
| $A_1$-TO | 378 – 380 |
| $E_1$-TO | 409 - 413 |
| $A_1$-LO | 574 - 579 |
| $E_1$-LO | 583 – 591 |
| $E_2^{low}$ | 99 – 101 |
| $E_2^{high}$ | 437 - 438 |

TABLE I

**FIGURE CAPTIONS**

FIG. 1. Un-polarized Raman spectrum of bulk ZnO, $\lambda_{exc}$ = 532nm. Raman active modes with the corresponding vibrations of the ions are indicated. The red arrows specify the dominant ions displacements.

FIG. 2. SEM images of a selection of AZO films deposited at increasing oxygen pressure (a) 2 Pa (b) 100 Pa (c) 160 Pa

FIG. 3. Variation of the lattice parameter c (a) and the vertical domain size (b) with oxygen pressure for ZnO (black) and AZO (red) films.

FIG. 4. Visible Raman spectra of ZnO (black) and AZO (red) films deposited at the same oxygen pressure (0.1, 2, 10 and 100 Pa), (a) $\lambda_{exc}$ = 532 nm, filter cut-off at 50 cm$^{-1}$; (b) $\lambda_{exc}$ = 457 nm, filter cut-off at 150 cm$^{-1}$. The reference positions of the modes are indicated with vertical lines (see fig. 1 and table I).

FIG. 5. Raman spectra of ZnO (black) and AZO (red) films deposited at the same oxygen pressure (0.1, 2, 10 and 100 Pa), $\lambda_{exc}$ = 325 nm, filter cut-off at 150 cm$^{-1}$. Vertical lines indicate the LO mode and two of its overtones. The increasing background is due to PL.

FIG. 6. Peak position (a) and width (b, c) of the two $E_2$ modes for ZnO (black) and AZO (red) films as a function of oxygen pressure. Data from Raman spectra at $\lambda_{exc}$ = 532 nm. Bulk values for peak position are indicated by horizontal lines.

FIG. 7. Intensity ratio between LO band and $E_2^{high}$ mode as a function of oxygen pressure. (a) comparison between ZnO (black) and AZO (red) films. Data from Raman spectra at $\lambda_{exc}$ = 532 nm. The horizontal line is set at the bulk value of 0,75 extracted the spectrum of bulk ZnO reported in fig.1. (b) Comparison of data extracted from Raman spectra at *green* and *blue* excitation for AZO film.

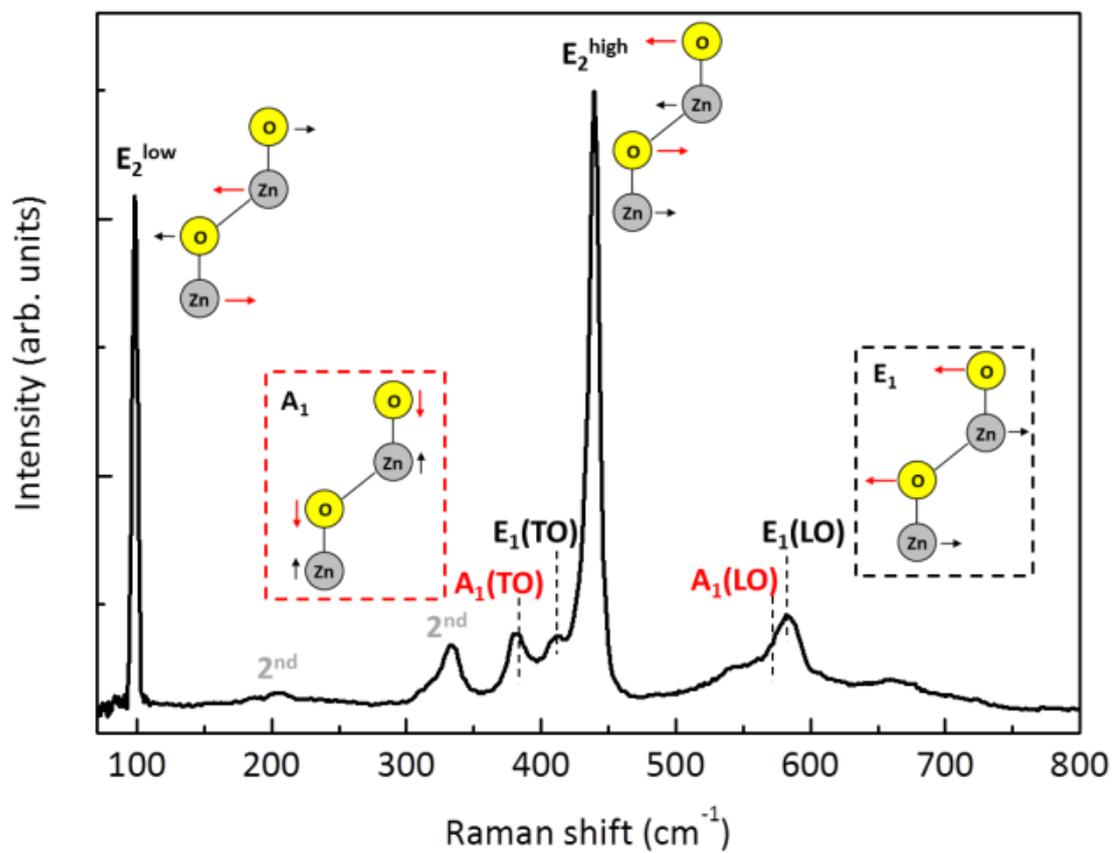

FIGURE 1

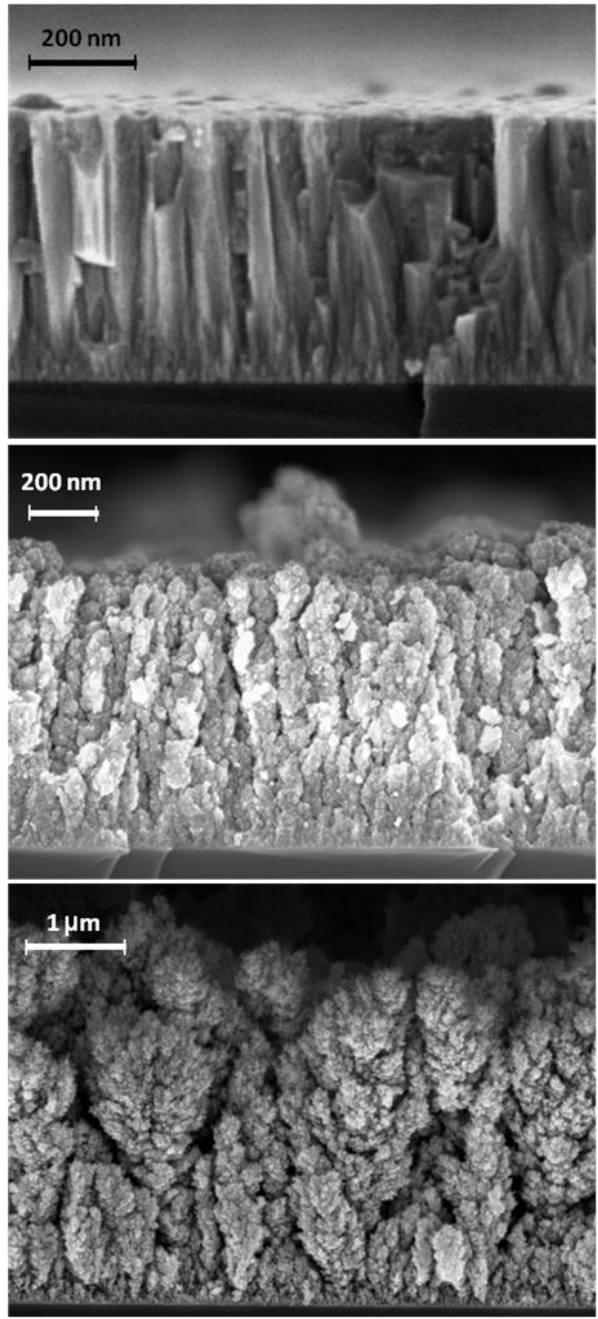

FIGURE 2

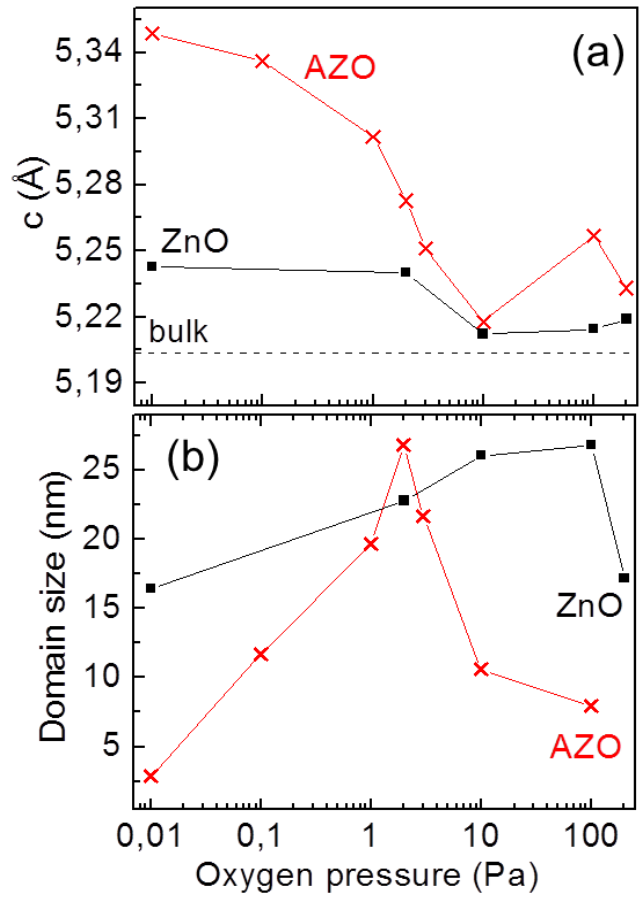

FIGURE 3

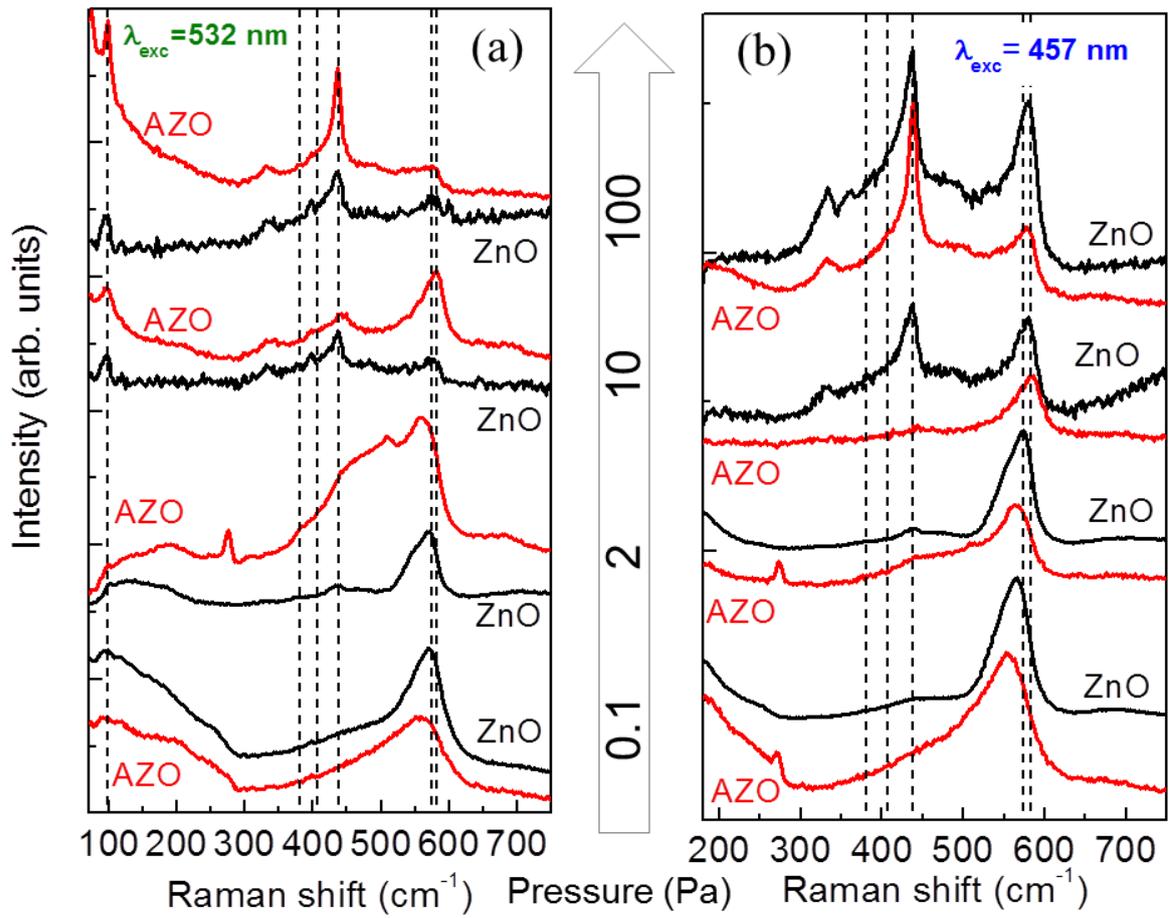

FIGURE 4

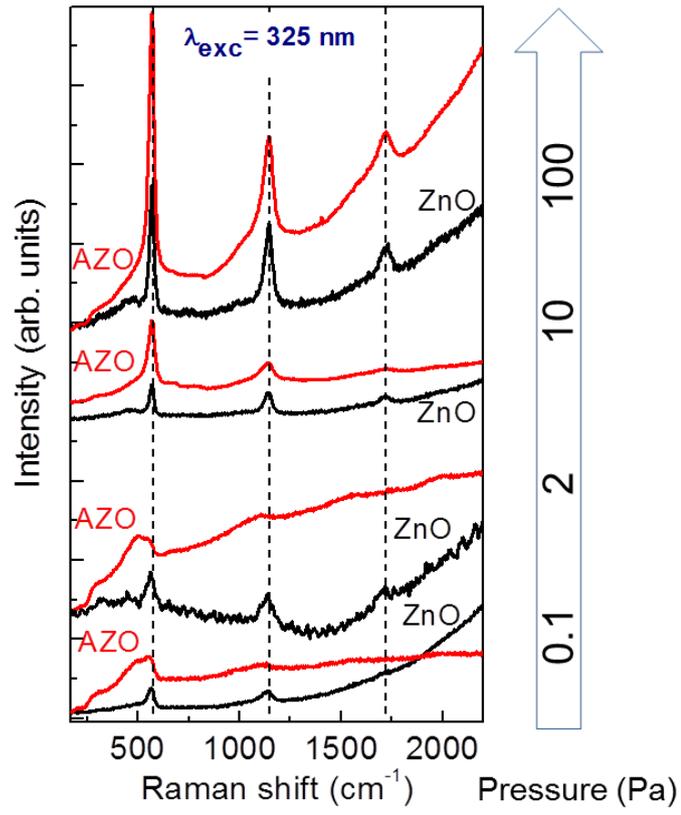

FIGURE 5

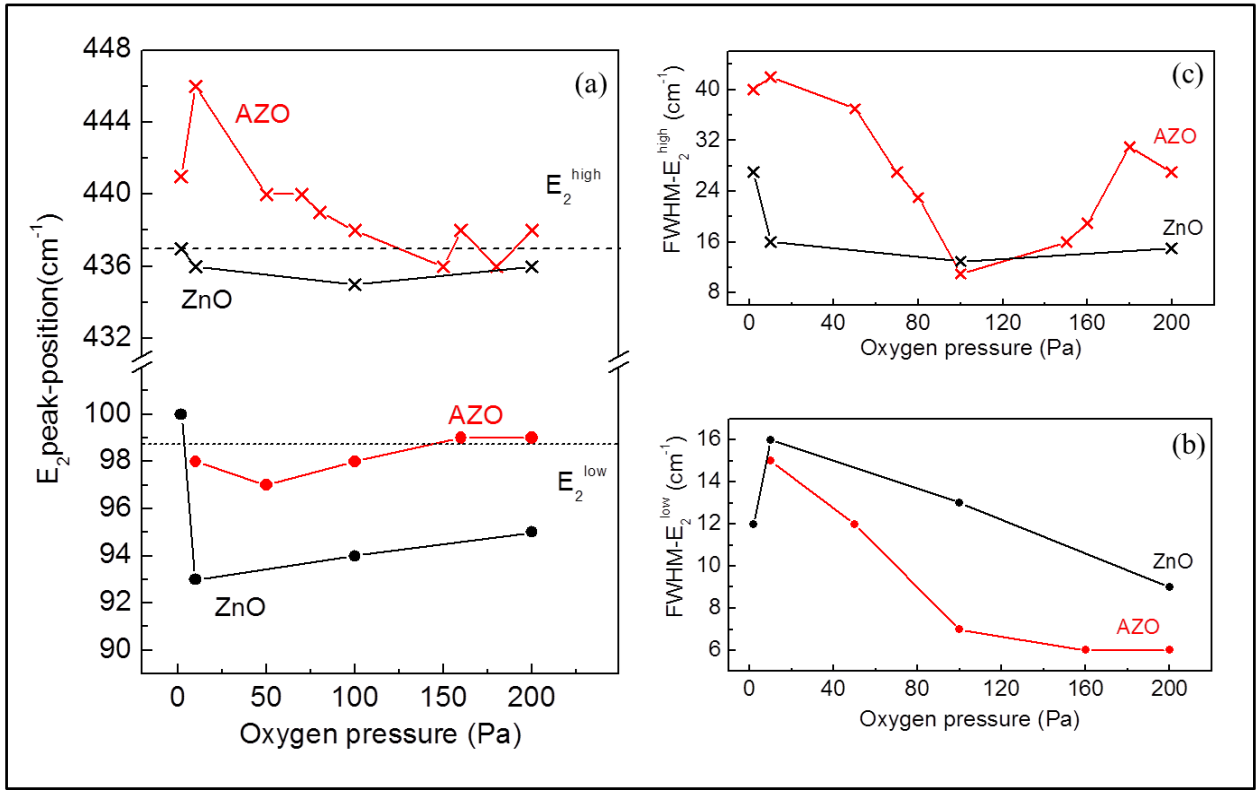

FIGURE 6

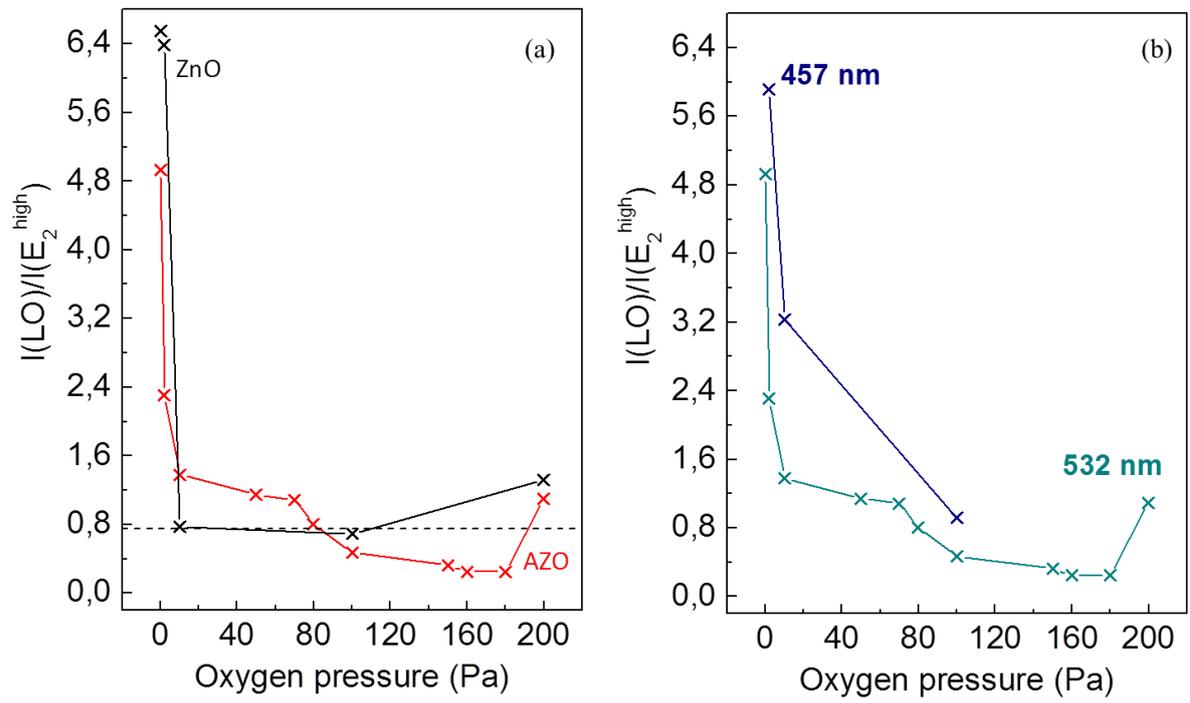

FIGURE 7